\documentstyle[12pt]{article}
\newcommand{\rf}[1]{(\ref{#1})}
\newcommand{\beq}{\begin{equation}}
\newcommand{\eeq}{\end{equation}}
\newcommand{\g}{\gamma}
\renewcommand{\l}{\lambda}

\newcommand{\n}{\nu}
\newcommand{\k}{\kappa}
\newcommand{\m}{\mu}

\newcommand{\sg}{\sigma}
\newcommand{\bea}{\begin{eqnarray}}
\newcommand{\eea}{\end{eqnarray}}

\newcommand{\cD}{{\cal D}}

\newcommand{\cT}{{\cal T}}

\newcommand{\bcT}{\bar{\cT}}
\newcommand{\bG}{\bar{G}}
\newcommand{\bmu}{\bar{\m}}
\newcommand{\bphi}{\bar{\phi}}
\newcommand{\bg}{\bar{\g}}
\newcommand{\Tr}{{\rm Tr}\;}

\renewcommand{\ni}{\noindent}

\newcommand{\oh}{\frac{1}{2}}

\newcommand{\vencm}{\vspace{1.0cm}}
\newcommand{\vs}{\vspace{0.2cm}}
\newcommand{\vhcm}{\vspace{0.5cm}}
\newcommand{\vtkcm}{\vspace{0.75cm}}

\begin{document}
\topmargin 0pt
\oddsidemargin 5mm
\headheight 0pt
\topskip 0mm

\addtolength{\baselineskip}{0.10\baselineskip}
\hfill NBI-HE-93-31

\hfill July 1993
\begin{center}

\vspace{36pt}
{ \bf    BARRIERS IN QUANTUM GRAVITY}
\footnote{Invited talk
presented at the ITCP spring school and workshop on string theory,
Trieste, Italy, April 1993. To appear in the proceedings.}
\end{center}

\vencm

\begin{center}
 J. AMBJ\O RN

{\it The Niels Bohr Institute\\
Blegdamsvej 17, DK-2100 Copenhagen \O , Denmark}
\end{center}

\vencm

\begin{center}
{ ABSTRACT}
\end{center}

\begin{center}
\begin{minipage}{12.7cm}
{\small

\ni
I discuss recent progress in our understanding of two barriers in
quantum gravity: $c > 1$ in the case of 2d quantum gravity and
$D > 2$ in the case of Euclidean Einstein-Hilbert gravity formulated
in space-time dimensions $D >2$.
}

\end{minipage}
\end{center}

\vencm

\ni 1.~~~~ {\bf c \mbox{\boldmath $  > $ } 1}

\vs

\ni
In the last four years there has been a tremendous progress in our
understanding of 2d quantum gravity and non-critical strings. However,
one of the main motivations for studying
non-critical strings was to be able to  formulate a string theory without
tachyons for dimensions of target space larger than one, and in this
respect there has been virtually no progress. Stated slightly differently:
we have still not understood  2d quantum gravity coupled to matter
with a central charge larger than one. If we use the discretized approach
\cite{david1,adf,kkm,david2,adfo} there seems to be no formal problems
with such a coupling. One has a statistical system of random
surfaces coupled to a number of gaussian scalar fields or coupled
to a number of spin systems, which only interact via the geometry
of the random surface. Formally the system is reflection positive and there
should be no tachyonic excitations in the system. However, two theorems
established soon after the invention of these models made it unlikely that
interesting theories would exist for $c > 1$. Recent results of numerical
nature seem to indicate loopholes related to the theorems
\cite{bh,hikami,ajt,adjt,wexler,bj,ckr}.
In the following I will review the theorems
and explain my present understanding of the loopholes.

\vhcm

\ni 1.1~~~ {\it The theorems}

\vs

\ni
The following theorem was proven in the context of 2d gravity coupled
to multiple Gaussian fields \cite{adf,adfo,adf1,ad}, but the proof can
presumably be generalized to other matter fields:
\newtheorem{theo}{theorem}
\begin{theo}
{}~~~~$\g \leq \oh$ ~~and~~ $\g > 0 ~\Rightarrow~ \g=\oh$.
\end{theo}
\begin{theo}
{}~~~~$\sg(\m_c) > 0$.
\end{theo}
In the first theorem $\g$ denotes the string susceptibility. In the
second theorem  $\sg(\m)$ denotes the string tension as a function of the
cosmological constant $\m$, and $\m_c$ is the critical value of the
cosmological constant, where the continuum limit is taken.
Theorem 2 is only relevant for multiple Gaussian fields since these can be
given the interpretation as target space coordinates of the string.
The string tension is defined by the exponential decay
of the one-loop function, the loop being
a fixed planar boundary $\partial A$
enclosing an area $A$. Let $G_\m(\partial A)$ denote the one-loop
Greens function. Then $\sg(\m)$ is defined by
\beq
G_\m(\partial A) \sim e^{-\sg(\m) A},~~~~~
\sg(\m) \equiv \lim_{A\to \infty} \frac{\log G_\m (\partial A)}{A}. \label{1}
\eeq

According the the KPZ formula $\g$ is an increasing
function of $c$, increasing to 0 as $c$ increases to 1,
while the formula leads to complex $\g$'s for $c >1$. It is natural
to interpret the first theorem as an indication that the random surfaces
degenerate to so-called branched polymers as soon as $c>1$, since
the value $\g=1/2$ is the generic value for branched polymer surfaces.
This interpretation is further corroborated by theorem 2 in the case of
multiple Gaussian fields provided we define the relation between
the bare (dimensionless) string tension $\sg(\m)$ and the ``continuum''
string tension $\sg_{phys}$ having dimension (mass)$^2$ in the usual way.
Let $a$ be a scaling parameter with dimension of length,
usually identified a typical lattice cut-off
in the regularized theory. We imagine that $a$ is a function of the
cosmological constant which scales to 0 for $\m \to \m_c$, and defined
with reference to some mass parameter of the theory which scales to
zero. One choice could e.g. be the mass
defined by the exponential decay of the two-point
function in target space (i.e. the ``tachyon'' mass in ordinary string
theory):
\beq
m(\m) \sim const. \cdot (\m-\m_c)^\n = m_{phys} a(\m). \label{2}
\eeq
If it is not possible to find a bare mass which scales to zero,
one would be tempted to conclude that the theory has no continuum
interpretation in target space.  If we {\it assume} the existence of
a bare mass which scales according to \rf{2} we get
$a(\m) \sim (\m-\m_c)^\n$,
and the physical mass $m_{phys}$ is simply {\it our}
choice of ratio between $m(\m)$ and $a(\m)$.
The relation between the  physical string tension
$\sg_{phys}$ and the bare string $\sg(\m)$ tension has to be defined
in a similar way:
\beq
\sg (\m)= \sg_{phys} a^2(\m).   \label{3}
\eeq
Since $a(\m)$ by assumption scales to zero we conclude that
$\sg_{phys} \to \infty$ for $\m \to \m_c$, due to theorem 2. The only
excitations of a surface with very large string tension are those
which do not increase minimal the area
dictated by the boundary conditions. The only possible outgrowths
will be thin tubes which are allowed to branch, i.e. precisely
branched polymer surfaces.

\vhcm

\ni 1.2~~~ {\it Theorem 1}

\vs

\ni
As mentioned above there has been accumulating evidence that especially
theorem 1 might be circumvented in models with various kind of matter
coupled to 2d quantum gravity. Let me mention possible loopholes
in the proof of this theorem\footnote{This analysis was first performed
by B. Durhuus in the case of multiple Ising spins, and most of the
following discussion of theorem 1 is inspired by discussions with him.}

Let us for notational simplicity consider pure 2d gravity. In that case
we know of course that $c=0$ and $\g = -1/2$ but we can apply the general
arguments to the model which is defined as follows:
\beq
Z(\m)= \sum_{T \in \cT}  e^{-\m |T|} \label{4}
\eeq
where $\cT$ denotes a suitable class of triangulations. We will here
restrict the topology to be spherical. Let us further define the $n$-puncture
function as\footnote{I order not to make the discussion too technical
we have chosen not to specify certain symmetry factors in eq. \rf{4} and eq.
\rf{5}.}
\beq
G_n (\m) = (-1)^n \frac{d^n}{d\m^n} Z(\m). \label{5}
\eeq
We have
\beq
G_n (\m) \geq G_2^n (\m),~~~~n >2.  \label{6}
\eeq
The geometrical interpretation of eq. \rf{6} is as follows:
The rhs of eq. \rf{6} can be viewed as a surface made out of
$n$ elongated balloons each with two marked points: one at the ``root'' and
one at the top of the balloon, all tied together at their ``roots'', this
common point being viewed as an interior point. In this way the rhs can
be viewed as a surface with $n$ marked points and
this restrict sum over surfaces with $n$
marked points is smaller that the sum over all surfaces with $n$ marked
points, i.e. the lhs of eq. \rf{6}. These arguments can be made rigorous
in  specific models (see for instance \cite{adf1}).

By definition of the string susceptibility we have
\beq
G_n(\m) \sim a.p. + (\m-\m_c)^{2-n-\g} + \cdots     \label{7}
\eeq
where $a.p.$ denotes an analytic part at $\m_c$ while $\cdots$ denote
less singular terms. From \rf{6} we get
\beq
\g < 1-\frac{1}{n-1}  \label{8}
\eeq
and the strongest bound on $\g$ comes from $n=3$ and is $\g \leq 1/2$.
It is tempting to use \rf{8} for $n=2$, in which case we would
get $\g \leq 0$. However, \rf{6} is not correct in that case due to a
non-unique decomposition of surfaces in a sequence of ``balloons''-like
surfaces.

We can derive an equation in the case where $n=2$ as follows: Let us
consider the one-point function. To be more precise we will consider
the function where one triangle is considered the external boundary.
This function has the same scaling behaviour as the one-point function
where only a vertex is kept fixed. Assume that the smallest allowed
loops in our class of triangulations are of length 3, i.e. are triangles.
Any 3-loop of links in the surface is either a triangle belonging to the
surface or it can be viewed as the boundary of a surface which grows out from
our the rest of the surface
and which can be separated from the rest
by cutting along the 3-loop. If we still use the notation ``one-point''
function for the surfaces with a triangle (a 3-loop)
as boundary we can write:
\beq
G_1(\m) = \sum_{T \in \bcT} e^{-\m |T|} \left( 1+G_1(\m)\right)^{|T|}
\equiv \sum_{T \in \bcT} e^{-\bmu |T|}
\label{9}
\eeq
where $\bcT$ denotes the class of triangulations which cannot be
separated in two by cutting along a loop of length 3. Clearly this
class of triangulations is a perfectly good one by its own right
and we can define a partition function and $n$-point functions
similar to \rf{4}-\rf{5}, only with $\bcT$ instead of $\cT$.
Let us denote the corresponding $n$-point functions with $\bG_n (\bmu)$,
where $\bmu$ denotes the cosmological constant
of the modified model. By definition we have
\beq
\bmu = \m-\log (1+G_1(\m)),~~~~~\bG_1 (\bmu) = G_1 (\m). \label{10}
\eeq
We can differentiate these equations with respect to $\m$:
\beq
\frac{d \bmu}{d \mu}= 1+ \phi(\m),~~~~~
\phi(\m)\equiv \frac{G_2(\m)}{1+G_1(\m)} \label{11}
\eeq
and
\beq
\phi(\m) = \frac{\bphi (\bmu)}{1-\bphi(\bmu)}. \label{12}
\eeq
$\phi$ is essentially the two-point function $G_2$. They have the
same critical behaviour and we will not distinguish between them.
$\bphi$ is defined as $\phi$, just using $\bG_1$, $\bG_2$ and $\bmu$ instead
of $G_1$, $G_2$ and $\m$. It can be viewed as the sum over surfaces
which connect the two boundary points (or boundary 3-loops) and where
the surfaces in the sum
have the property that they cannot be separated in
two components, each of which contains one of the boundaries,
by a cut along a 3-loop. By expanding the denominator
eq. \rf{12} gets the obvious
graphical interpretation of the two-point function $\phi(\m)$
as the sum over all surfaces (of trivial topology)
made up by successive gluing of ``irreducible'' $\bphi$-surfaces along
their boundaries, such that only two boundaries are left as external.

If we assume that $\g > 0$, we know from \rf{7} that $\phi(\m)$ will diverge
for $\m \to \m_c$. From \rf{12} this is impossible for $\bphi(\bmu(\m))$
since it follows that
\beq
\bphi(\bmu_0)= 1,~~~~~\bmu_0 \equiv \bmu(\m_c)~~(\geq \bmu_c). \label{13}
\eeq
We conclude that either $\bmu_0 > \bmu_c$, the critical point for the
modified model defined by the class of triangulations $\bcT$, or $\bg$
corresponding to $\bcT$ has to  satisfy $\bg \leq 0$ in order that
$\bphi(\bmu_c)$ is bounded (in fact $\bphi(\bmu_c)=1$ by \rf{12}).

In the first case we can expand $\bphi(\bmu)$ in a Taylor series around
$\bmu_0$ and use that due to \rf{11} we have
\beq
\bmu_0-\bmu \approx \left.\frac{d\bmu}{d\m}\right|_{\m}(\m_c-\m) \approx
-(\m-\m_c)^{1-\g} \label{14}
\eeq
\beq
\bphi(\bmu) \approx 1+\bphi'(\bmu_0)(\bmu-\bmu_0)
            \approx 1+\bphi'(\bmu_0) (\m-\m_c)^{1-\g}. \label{14a}
\eeq
If we insert $\phi(\m) \approx (\m-\m_c)^{-\g}$ and \rf{14a} for $\bphi$ in
eq. \rf{12} we get the promised result $\g=1/2$, i.e. the exponent
corresponding to branched polymers.

In the Taylor expansion around $\bmu_0$ we used that $\bphi'(\bmu_0) < 0$.
This follows from the fact that the two-point function is the sum over
surfaces with positive weight $\exp(-\bmu |T|)$. If we however lift the
restriction that the surfaces should enter with positive weight, by adding
to the definition \rf{4} a weight $\rho(T)$ which can take positive
and negative values according to the triangulation, it is possible to
arrange that
\beq
\bphi'(\bmu_0)=\cdots = \bphi^{(n-1)}(\bmu_0) =0,~~~~
\bphi^{(n)}(\bmu_0)\neq 0.  \label{15}
\eeq
In this case we get instead of eq. \rf{14a}
\beq
\bphi(\bmu) \approx 1+\bphi^{(n)}(\bmu_0)(\bmu-\bmu_0)^n
            \approx 1+\bphi^{(n)}(\bmu_0) (\m-\m_c)^{(1-\g)n}. \label{16}
\eeq
If we insert this in eq. \rf{12} we get
\beq
\g = 1-\frac{1}{n+1}. \label{17}
\eeq
We can clearly arrange that $\g > 1/2$. But it is unlikely that the theory
is different from a theory of branched polymers. In fact there exists a
theory of so-called multicritical branched polymers, where the probability
for branching can be negative \cite{adj}. The possible values of $\g$
for such branched polymers are precisely given by \rf{17}.

The only possibility to get something non-trivial for $\g > 0$ seems
to be the situation where $\bmu_0 = \bmu_c$. In this case we
know that $\bg \neq \g$. In fact we have to have $\bg \leq 0$ since
$\bphi(\bmu_c) =1$. We can no longer make a Taylor expansion as in \rf{14a}.
We can however use \rf{14} and the fact
that $\bphi(\bmu)$ (at least for $\bg \geq -1$) by the definition of $\bg$
behaves like
\beq
\bphi(\bmu) \approx 1 + (\bmu-\bmu_c)^{-\bg}
\approx 1 + (\m-\m_c)^{-\bg (1-\g)}. \label{18}
\eeq
If we insert this in \rf{12} we get
\beq
\g = - \frac{\bg}{1-\bg},~~~~~~\bg=-\frac{1}{n} ~\Rightarrow~
\g = \frac{1}{n+1}  \label{19}
\eeq
which gives a remarkable relation between $\g$ and $\bg$, first obtained
by Durhuus \cite{durhuus}. The interpretation of this relation is
however a little disappointing. We see that it might be possible
to obtain a $\g$ between 0 and $1/2$. However, it does not really
reflect a new theory. It seems rather to be a ``polymer-like'' or
``bubble-like'' manifestation of an underlying $c < 0$ theory.

The above analysis only opened for the possibility that $0 < \g < 1/2$.
It did not provide any examples. The treatment was also simplified in the
sense that we used the notation of pure gravity where we know by explicit
calculation that $\g < 0$. There is an example of a model which realizes
precisely the scenario mentioned above, a modified matrix model for
generating random surfaces, first introduced in \cite{bombay} and analysed
further in \cite{komt,alvarez}. The partition function (in the sense of
generating connected random surfaces) is given by
\beq
Z(\m,g) = \log \left\{ \int dM \exp\left[-\m N \Tr V(M) +
g \m^2 (\Tr M^2)^2\right] \right\}. \label{20}
\eeq
where $M$ denotes an $N \times N$ Hermitian matrix and $V(M)$ is a potential
corresponding the $n^{th}$ multicritical matrix model. If $g=0$ this
matrix model will generate random surfaces with $\g = -1/n$. We have
a critical
line in the $(\m,g)$-coupling constant plane and at a certain point along
this line the critical behaviour changes and $\g$ jumps from $-1/n$ to
$1/(n+1)$. The geometrical interpretation of the model is that the
term $(\Tr M^2)^2$ introduces a new kind of vertex in the model, such
that to leading order in $1/N$ ordinary spherical surfaces dominates, except
that two (or more) such surfaces can touch each other by means of the
new vertices, the topology of the total surface still being spherical.
The jump in $\g$ is related to a change in behaviour the effective
``touching'' coupling constant $N\bar{g} = -g\m <\Tr M^2>$ as a function
of $\m$, from being analytic to non-analytic, i.e. a situation
precisely as described above in general terms.

The other class of models where one could imagine that $0 < \g < 1/2$ is
multiple Ising models coupled to gravity. Durhuus has analysed this
situation along the lines indicated above\cite{durhuus}. Things are
slightly more complicated since we will now have several coupling
constants and the scalar equations written above will be replaced by
matrix equations. However, the result is unchanged: The only possible
alternative to $\g=1/2$ seems to be that $\g = 1/(n+1)$, and this model is
somehow a shadow of an underlying unitary model with $\bg = -1/n$ and
a central charge $\bar{c} < 1$.

There is some numerical evidence in favor of such a scenario \cite{ajt,adjt},
but since it is now more clear what to look after it should be possible to
provide much better numerical verification of the above scenario for
models with $c >1$.

\vhcm

\ni 1.3~~~{\it Theorem 2}

\vs

\ni
Theorem 2 is more specific for discretized string models, and the
loopholes not obvious, if any. One cure, which is inspired by the
geometrical picture of spiky polymers emerging from a minimal surface,
attempts to add by hand extrinsic curvature terms.
The model was first formulated in \cite{adj1,adfj}, partly inspired
by theorem 2 (\cite{ad}), partly by the fact that one could get
extrinsic curvature terms by integrating out the fermionic part
of the superstring. It seems difficult to obtain analytic results for
the model (see however \cite{adfj1} for a renormalization group approach),
and most work has been of numerical nature. From large scale Monte Carlo
simulations the following have been established (with the usual reservation
of numerical simulations, which have a lot in common with experimental
physics, including the ability to produce results which have to be retracted
eventually):  As a function of the extrinsic curvature coupling constant
$\l$ we have two phases : One phase where the physics is the same as
for  $\l =0$, i.e. the string tension $\sg(\m,\l)$ is not scaling
to zero as $\m \to \m_c(\l)$ (the critical value of the cosmological
constant $\m$ now depends on $\l$), and , for $\l > \l_c$, a now phase
where the surfaces essentially are ``smooth'' embeddings in target space
with Hausdorff dimension $d_H =2$. {\it At the transition point} $\l_c$
we have according to the numerical simulations \cite{catterall,aijpv,aijp}
that the string tension scales to zero as $\m \to \m_c(\l_c)$. In addition
the mass gap $m(\m,\l_c)$ of the two-point function scales to zero for
$\m \to \m_c(\l_c)$ in such a way that
\beq
\frac{m^2(\m,\l_c)}{\sg(\m,\l_c)} \approx constant. \label{21}
\eeq
According to the discussion above this is precisely what is needed
in order to achieve an interesting continuum limit which allow the
interpretation as a string theory. It is, however, at present unclear
which  continuum theory could serve as a candidate for the
continuum limit of the regularized string theory.

It is interesting to view the model with extrinsic curvature in the framework
of conformal field theory coupled to 2d quantum gravity. From this
point of view we should first consider a fixed regular triangulation.
The lattice points are mapped to target space (say $R^3$, where most
of the numerical simulations are actually performed) and the action
is the usual Gaussian action plus extrinsic curvature (for details
we refer i.e. to \cite{adj2,wheater}). It is local, but non-polynomial in
the target space coordinates $x_i$, due to the extrinsic curvature
term. The model has also attracted the attention of the
solid state physicists and it is known that it has an infrared fixed
point at $\l=0$ and a ultraviolet fixed point at a certain critical
value of $\l$. The central charge for $\l=0$ is 3, since we in that
case  have three free gaussian fields. The central charge corresponding
to the ultraviolet fixed point is unknown. In case of a unitary theory
one would (by the $c$-theorem) expect $c \geq 3$, but it is not
known if the conformal field theory corresponding to the ultraviolet
fixed point is unitary. Coupling to gravity should now, according to the
standard procedure, consist in replacing the regular triangulation
with  dynamical triangulations. The phase transition
reported above for dynamical triangulations
is precisely the one analogue of the transition corresponding to the UV-fixed
point for the regular triangulation. Have we manage to couple a theory
with $c > 1$ to quantum gravity with a non-trivial result?
Certainly this scenario deserves further attention.

\vtkcm

\ni 2.~~~~{\bf QUANTUM GRAVITY FOR D \mbox{\boldmath $ >$} 2}

\vs

\ni Attempts to quantize quantum gravity in dimensions $D > 2$ differ
radically from the attempts described above for $D=2$. For $D >2$ we
have not a well understood theory when the rotation to Euclidean space
is performed: Due to the conformal mode the Einstein-Hilbert action
is unbounded from below. The Euclidean path integral is not well defined.
One way to avoid this problem was suggested by Hawking: An analytical
continuation of the conformal mode. But it is not known if
the method will work beyond perturbation theory (see \cite{ag} for an
example where analytic continuation works to all order in perturbation theory
but fails non-perturbatively). Another suggestion is to use
stochastic quantization \cite{gh}. Also this method has the flavor
of being somewhat arbitrary. In addition to the problems with the
conformal mode we are in quantum gravity faced with the question
of topology. Should all topologies be included in the quantum theory,
and if this is the case have can we imagine this should be done, since
it is impossible  to classify 4-geometries?

Contrary to two dimensions we are obviously
in the situation where we try to define in a non-perturbative way a
theory, which existence can be  said to be somewhat doubtful. The only reason
we engage in such a project is the fact that we know from
experimental evidence that there exists {\it a theory
of gravity}, which presumably should be quantized. One candidate for a
non-perturbative definition is of course string theory, but since
it is presently by no means obvious that string theory
itself exists in a non-perturbative formulation, one can seriously
doubt that it can be used  define quantum gravity in a non-perturbative
way.

Simplicial quantum gravity is an attempt to give a non-perturbative
definition of Euclidean gravity. Its main advantage is that it has
been shown to work well (at least as well as any method based on
a continuum formulation) in $D=2$. In addition it provides a very
natural geometric discretization of the Einstein-Hilbert action.
A drawback is that it deals in a superficial way
with the conformal mode problem\footnote{This is however true for
almost all regularizations.} since the regularization is such that
for a fixed lattice volume the action is bounded from below.
Of course this just postpone the conformal mode problem until the
scaling limit has to be taken. In addition it has little to say about
the summation over topologies, but at least it highlights the problem,
since the model is simply not well defined unless we restrict the topology.
In the following we  will always assume the simplest topology, i.e. $S^4$
in $D=4$.

\vhcm

\ni 2.1~~~{\it The model}

\vs

\ni As in $D=2$ we imagine the manifolds  which enters in the regularized
path integral to be constructed by gluing together identical simplexes,
in this case 4-simplexes rather than
the equilateral triangles used in $D=2$.
The curvature of such a piecewise linear manifold is given by Regge
calculus, which in this case becomes extremely simple (see for instance
\cite{ajk,abjk} and reference herein): Suppose that our manifold
is constructed out of $N_4$ simplexes and that it has $N_2$ two-dimensional
sub-simplexes. Using the construction of Regge a little algebra leads to
to the following identifications (which we write down in $D$ dimensions
for the sake of generality):
\beq
\int d^D \xi \sqrt{g} ~\propto~ N_D \label{22}
\eeq
\beq
\int d^D \xi \sqrt{g} R ~\propto~ \frac{2\pi}{\arccos (1/D)} N_{D-2}-
\frac{D(D+1)}{2} N_D. \label{23}
\eeq
For a given four-dimensional triangulation $T$ we can therefore write the
cosmological term plus the Einstein-Hilbert action as
\beq
S[T] = \k_4 N_4(T) -\k_2 N_2(T) \label{24}
\eeq
where $\k_2$ is proportional to the inverse of the bare gravitational
coupling constant and $\k_4$ involves a combination of the bare cosmological
coupling constant and the bare gravitational coupling constant.

The formal recipe for going from the continuum functional integral to the
discretized one is now:
\beq
\int \cD[g_{\m\n}] ~~\to~~ \sum_{T \in \cT} \label{25}
\eeq
\beq
\int \cD[g_{\m\n}] e^{-S[g]}~~\to~~
\sum_{T \in \cT} e^{-S[T]} \label{26}
\eeq
where $[g_{\m\n}]$ denotes equivalence classes of metrics, i.e. the volume
of the diffeomorphism group is divided out, and it
is understood that the functional integral is only over connected
four-manifolds of topology $S^4$.

It is straight forward to generalized the action to include the coupling to
matter, e.g. spin systems or Gaussian fields (we refer to \cite{abjk1} for
details).

The rhs of eq. \rf{26} shares a number of properties with its
two-dimensional analogue: The number of triangulations of topology $S^4$
and a given volume $N_4$ is exponentially bounded\footnote{This is only
verified by numerical methods. It would be interesting to have an analytic
proof of this.}. This implies that the partition function for a given
$\k_2$ has a critical $\k_4$ where the infinite volume limit can
be taken. It does not imply that we necessarily have an interesting
continuum theory. One has to scan $\k_2$ to look for divergent correlation
lengths, at least if we want to use the standard intuition from
statistical mechanics and the theory of critical phenomena. Unfortunately
the only tool which seems useful at the moment in the analysis of the
scaling limit is numerical simulations.

\vhcm

\ni 2.2~~~{\it Observables and numerical simulations}

\vs

\ni
Due to the invariance  under diffeomorphism and the fact that
we in quantum gravity have to integrate over all Riemannian manifolds,
the observables which are most readily available are averages of invariant
local operators like the curvature $R(x)$. A non-local observable
is the integrated curvature-curvature correlation.
In a continuum formulation it would be
\beq
 \chi (\k_{2}) \equiv \langle \int d^4 \xi_1 \;
 d^4 \xi_2 \; \sqrt{g(\xi_1)} R(\xi_1)
\;\sqrt{g(\xi_2)} R(\xi_2) \;\rangle-
\langle\int d^4 \xi \sqrt{g(\xi)}R(\xi)\rangle^2. \label{27}
\eeq
In a lattice regularized theory one would expect that away from the critical
points short range fluctuations will prevail, while approaching
the critical point long range fluctuation might be important and would
result in an increase of $\chi (\k_{2})$. The observable
$\chi (\k_{2})$ is the second
derivative of the free energy $F= -\ln Z$ with respect to the
inverse gravitational coupling constant.
In the case where the volume $N_{4}$ is kept fixed we have
\beq
\chi (\k_{2},N_{4}) \sim
\langle N_{2}^2\rangle_{N_{4}}-\langle N_{2}\rangle_{N_4}^2
{}~=~ - \frac{d^2 \ln Z(\k_{2},N_{4})}{d \k_{2}^2} label{28}
\eeq
{}From the above discussion it follows that we have to search for values
of $\k_2$  where $\chi (\k_{2},N_{4})/N_{4}$
diverges in the infinite volume limit $N_{4} \to \infty$.

An independent susceptibility is  associated with volume
fluctuations:
\beq
\chi_V (\k_{2},\k_{4}) \sim
\langle N_{4}^2\rangle-\langle N_{4}\rangle^2
{}~=~ - \frac{d^2 \ln Z(\k_{2},\k_{4})}{d \k_{4}^2} \label{29}
\eeq
{\it Assume} that $Z(\k_{2},N_4)$ has the form
\beq
Z(\k_{2},N_4) \sim N_4^{\g(\k_{2})-3} e^{\k_4^c(\k_{2}) N_4}\;
\left( 1 + O(1/N_4) \right).       \label{30}
\eeq
This is the case for $D=2$. For higher $D$ it is of course
a necessity for the existence of the model that $Z(\k_{2},N_4)$
is exponentially bounded, but it is by no means clear that subleading
corrections should appear as a power-law correction to
the exponential factor. If it nevertheless does, we can identify
$\g(\k_{2})$ with the  critical exponent for the volume fluctuations
at the critical point $\k_4^c (\k_{2})$:
\beq
\chi_V (\k_{2},\k_{4}) \sim
\frac{1}{(\k_{4}-\k_{4}^c(\k_{2}))^{\gamma(\k_{2})}} \label{31}
\eeq
for $\k_4$ close to $\k_4^c$.

Another observable is the Hausdorff dimension. One can define the
Hausdorff dimension in a number of ways, which are not necessarily
equivalent. Here we will simply measure the average volume
$V(r)$ contained within a radius $r$ from a given point.
In \cite{adj3} the concept of a {\it cosmological Hausdorff
dimension} $d_{ch}$ was defined. It essentially denotes the power
which relates the average radius of the ensemble of universes of
a fixed volume to this volume:
\beq
\langle {\rm Radius}\rangle_{N_{4}}~ \sim~ N_{4}^{1/d_{ch}}. \label{32}
\eeq
{}From the distribution $V(r)$ we can try to extract $d_{ch}$. If
 for large $r$ we have the behaviour
\beq
V(r)~ \sim~ r^{d_h}   \label{33}
\eeq
we can identify $d_{ch}$ and $d_h$.

By the use of Regge calculus it is straightforward to convert these
continuum formulas to our piecewise flat manifolds. Let me now summarize
some of the numerical results obtained with the last $1\oh$ year by
various groups \cite{aj,am,am1,ajk,bm}.

The measurements of $\chi (\k_2)$ show a peak at $\k_2 \approx 1.1$.
We will denote this value $\k_2^c$.
At this point there indeed seems to be some kind of phase transition
in geometry. For $\k_2   < \k_2^c$ the radius of the typical universes
generated by computer is very small, and the radius grows very slowly with
the volume $N_4$, indicating a large fractal dimension of space-time.
For $\k_2 > \k_2^c$ the situation is the opposite: The radius of the
typical universe is large, in fact much larger than one would naively
expect, and there is a strong dependence on the volume $N_4$. A closer
analysis of the distribution of distances reveals that the
geometrical structures generated are of low fractal dimension
(between 1 and 2). If we approach  $k_2^c$ it seems that we have some
chance of generate structures with a fractal dimension which is not
that different from 4, but it is difficult to measure the Hausdorff
dimension precisely. At a first glance $\k_2^c$ seems to be the
obvious candidate for a critical point where we can take the continuum
limit. There is one obstacle, however. It turns out that the scalar
curvature does not scale at the transition point. Since the continuum
curvature has the dimensions of $({\rm mass})^2$,  standard scaling
arguments will tell us that that the bare curvature has to scale
to zero at the critical point. This seems not to be the case
in the numerical simulations\footnote{This problem may be similar
in nature to the non-scaling of the string tension mentioned above.}
and the manifolds which are generated
can therefore not in any simple way be identified with smooth
continuum manifolds!

In order to make contact to various continuum models for
quantum gravity it is important to measure critical
exponents which can be calculated in  a continuum approach.
One such exponent is  $\g(\k_2)$, the exponent of volume fluctuations,
defined by eq. \rf{31}. A convenient way of performing the measurement
of $\g(\k_2)$ is by counting so-called baby universes \cite{ajjk,ajt}.
These are just parts of the triangulated piecewise linear manifold
which can be separated from the rest by cutting the manifold along
a closed, minimal three-dimensional submanifold, i.e. the hyper-surface
of a 4-simplex. In most cases such a three-manifold {\it will} just
be the hyper-surface of a 4-simplex which belong to the triangulation.
But there exists situations where the ``interior'' of the hyper-surface
does not belong to the triangulation\footnote{The two-dimensional analogue
is to cut a surface along a 3-loop. In most cases one will just cut off
a triangle, but if the three loop is a ``bottleneck'' separating two
``blobs'' of surfaces we have a genuine baby-universe situation.} and
in this case the universe will be separated in two smaller universes of
volume $B$ and $N_4-B$, where $N_4$ is the original volume of the
universe and we assume $B < N_4/2$. We call $B$ the volume of the baby
universe and it can be shown that under the assumption of a distribution
like \rf{30}, the distribution of baby universes  will be
\beq
n(B) \sim N_4 B^{\g(\k_2)-2},~~~~B << N_4. \label{34}
\eeq

The results of measuring the distribution of baby universes indicate
that the assumption of a distribution like \rf{30} fails\footnote{In
\cite{ajjk} we extract $\g$'s for $\k_2 < \k_2^c$, but it appears
that the baby universe volumes used were too small to do it in a
reliable way.}
for $\k_2 < k_2^c$ but is fulfilled for $k_2 > k_2^c$.
This is in agreement with similar observations made for
simplicial quantum gravity in $D=3$\cite{av,bk,av1,abkv}.
The value extracted close to $\k_2^c$ is approximately $-0.5\pm 0.2$,
i.e. close to the value for two-dimensional gravity!

There are various continuum candidate theories with which we can
hope to compare the numerical simulations. One candidate is the
restricted theory of self-dual conformal gravity. This theory
allows an analysis by 2d conformal techniques \cite{smidhuber}
and an ultraviolet fixed point is found around which $\g$ can be
calculated ($\g = -0.67$) and it is found to agree within errorbars
with the value measured at  $\k_2^c$. Another model which is
somewhat similar in spirit to the self-dual conformal gravity
in the sense that the physics of the conformal mode is emphasized
is a mini-super space model described in \cite{amm}. It does not
yet predict a value of $\g$, but the structure of the phase diagram
has some similarity to the one found in simplicial quantum gravity.
Finally a model of quantum gravity based on the $1/\epsilon$ expansion
around $D=2$ might be of interest as a regularized model which allow
us to compute critical exponents\cite{kn,kkn} and a comparison with
the $\g(\k_2)$'s determined by numerical simulations will be an important
task for the future.

\vtkcm

\ni {\bf Acknowledgement} It is a pleasure to thank Bergfinnur Durhuus
for explaining to me his ideas about $c >1$ Ising spin models. They
formed the foundation for section 1.2.

\addtolength{\baselineskip}{-0.30\baselineskip}


\begin{thebibliography}{99}
\bibitem{david1}F. David, Nucl. Phys. {\bf B257} (1985) 45.
\bibitem{adf}J. Ambj\o rn, B. Durhuus and J. Fr\"{o}hlich, Nucl. Phys.
{\bf B257}
 (1985) 433.
\bibitem{david2}F. David, Nucl. Phys. {\bf B257} (1985) 543.
\bibitem{kkm}V. A. Kazakov, I. K. Kostov and A. . Migdal,
Phys. Lett. {\bf 157B}
 (1985) 295.
\bibitem{adfo}J. Ambj\o rn, B. Durhuus, J. Fr\"{o}hlich and P. Orland,
Nucl.Phys. {\bf B270} (1986) 457.
\bibitem{bh}E. Brezin and  S. Hikami, Phys.Lett. {\bf B295} (1992) 209.
\bibitem{hikami}S.  Hikami, Phys.Lett. {\bf B305} (1993) 327.
\bibitem{ajt}J. Ambj\o rn, S. Jain and G. Thorleifsson, Phys.Lett.
{\bf B307} (1993) 34.
\bibitem{adjt}J. Ambj\o rn, B. Durhuus, T. Jonsson and G. Thorleifsson,
Nucl.Phys. {\bf B398} (1993) 568.
\bibitem{wexler}M. Wexler, {\it Matrix models on large graphs.},
PUPT-1398,
Bulletin Board: hep-th@xxx.lanl.gov - 9305041;
{\it Low temperature expansion of matrix models.},
PUPT-1384,
Bulletin Board: hep-th@xxx.lanl.gov - 9303146

\bibitem{bj} C.A. baillie and J.D. Johnston, Phys.Lett. {\bf B286} (1992) 44.
\bibitem{ckr}S. M Catterall, J.B. Kogut and R.L. Renken,
Phys.Lett. {\bf B292} (1992) 277.
\bibitem{adf1}J. Ambj\o rn, B. Durhuus and J. Fr\"{o}hlich,
 Nucl.Phys. {\bf B275} [FS17] (1986) 161.
\bibitem{adj}J. Ambj\o rn, B. Durhuus, T. Jonsson,
 Phys.Lett. {\bf B244} (1990) 403
\bibitem{durhuus} B. Durhuus, unpublished.

\bibitem{bombay}y S.R. Das, A. Dhar, A.M. Sengupta and S.R. Wadia,
Mod.Phys.Lett. {\bf A5} (1990) 1041.
\bibitem{komt}y G.P. KorchemskyPhys.Lett. {\bf B296} (1992) 323;
{\it Matrix models perturbed by higher curvature terms}, UPRF-92-334.

\bibitem{alvarez}y L. Alvarez-Gaume, J.L.F. Barbon, and C. Crnkovic,
Nucl.Phys. {\bf B394} (1993) 383.

\bibitem{adj1}J. Ambj\o rn, B. Durhuus, T. Jonsson,
Phys.Rev.Lett. {\bf58} (1987) 2619.
\bibitem{adfj}J. Ambj\o rn, B. Durhuus, J. Fr\"{o}hlich and T. Jonsson,
              Nucl.Phys. {\bf B290 [FS20]}  (1987) 480.
\bibitem{ad}J. Ambj\o rn and B. Durhuus, Phys.Lett {\bf 188B}  (1987) 253.

\bibitem{adfj1}J. Ambj\o rn, B. Durhuus, J. Fr\"{o}hlich and T. Jonsson,
J. Stat. Phys. {\bf 55} (1989) 29.

\bibitem{catterall}S.M.~Catterall, Phys.Lett {\bf B243} (1990) 121.
\bibitem{aijpv}J. Ambj\o rn, A. Irb\"{a}ck, J. Jurkiewicz  and B.
 Petersson, Nucl. Phys. {\bf B393} (1993) 571.
\bibitem{aijp}J. Ambj\o rn, A. Irb\"{a}ck, J. Jurkiewicz, B.
 Petersson and  S. Varsted, Phys.Lett {\bf B275} (1992) 295.
\bibitem{adj2}J. Ambj\o rn, B. Durhuus, T. Jonsson,
Nucl. Phys. {\bf B316} (1989) 526.
\bibitem{wheater}R. G. Harnish and J.F. Wheater, Nucl.Phys. {\bf B350}
 (1991) 861.
\bibitem{gh}J. Greensite and M. Halpern, Nucl.Phys. {\bf B242} (1984) 167.
\bibitem{ag}J. Ambj\o rn and J. Greensite,
 Phys.Lett. {\bf B254} (1991) 66.

\bibitem{ajk}J. Ambj\o rn, J. Jurkiewicz and C.F. Kristjansen,
Nucl.Phys. {\bf B393} (1993) 601.
\bibitem{abjk}J. Ambj\o rn, Z. Burda, J. Jurkiewicz and C.F.
Kristjansen, Acta Physica Polonica {\bf B23} (1992) 991.

\bibitem{abjk1}J. Ambj\o rn, Z. Burda, J. Jurkiewicz and C.F.
Kristjansen, {\it 4d quantum gravity coupled to matter}
NBI-HE-93-3, January 1993.

\bibitem{adj3}J. Ambj\o rn, B. Durhuus, T. Jonsson,
Mod.Phys.Lett. {\bf A6} (1991) 1133.

\bibitem{aj}J. Ambj\o rn and J. Jurkiewicz, Phys.Lett
{\bf B278} (1992) 42.
\bibitem{am}M.E. Agishtein and  A.A. Migdal, Nucl.Phys. {\bf B385}
(1992) 395.
\bibitem{am1}M.E. Agishtein and A.A. Migdal
Mod.Phys.Lett. {\bf A7} (1992) 1039.
\bibitem{ajjk} J. Ambj\o rn, S. Jain, J. Jurkiewicz and C.F. Kristjansen,
Phys.Lett. {\bf B305} (1993) 208.
\bibitem{bm}B. Br\"{u}gmann, {\it Non-uniform Measure in
Four-Dimensional simplicial Quamtum Gravity}, SU-GP-92/9-1;
B. Br\"{u}gmann and E. Marinari, {\it 4d Simplicial
Quantum Gravity with a Non-Trivial Measure}, SU-GP-92/9-2.

\bibitem{av} J. Ambj\o rn and S. Varsted, Phys. Lett. {\bf B226}, (1991) 285.
\bibitem{bk}D. Boulatov and A. Krzywicki, Mod.Phys.Lett. {\bf A6}
(1991) 3005.
\bibitem{av1}J. Ambj\o rn and S. Varsted, Nucl.Phys. {\bf B373} (1992) 557.
\bibitem{abkv}J. Ambj\o rn, D. Boulatov, A. Krzywicki and S. Varsted,
Phys.Lett {\bf B276} (1992) 432.

\bibitem{smidhuber} C. Schmidhuber
Nucl.Phys. {\bf B390} (1993) 188.

\bibitem{amm}   I. Antoniadis, P.O. Mazur and  E. Mottola,
{\it Scaling behavior of quantum four - geometries.}
CPTH-A214-1292. Bulletin Board: hep-th@xxx.lanl.gov - 9301002

\bibitem{kn}H. Kawai and M. Ninomiya,
Nucl.Phys. {\bf B336} (1990) 115.

\bibitem{kkn} H. Kawai, Y. Kitazawa and  M. Ninomiya
Nucl.Phys. {\bf B393} (1993) 280; {\it  Ultraviolet
stable fixed point and scaling relations in
(2+epsilon)-dimensional quantum gravity.},
UT-636. Bulletin Board: hep-th@xxx.lanl.gov - 9303123

\end{thebibliography}
\end{document}